# A Global Prospective of the Indian Optical and Near-Infrared Observational Facilities in the Field of Astronomy and Astrophysics: a review

**Ram Sagar**
**Indian Institute of Astrophysics, Sarajapur Road, Bangalore 560034 and**
**Aryabhatta Research Institute of Observational Sciences (ARIES), Manora Peak, Nainital, 263 002**

**Corresponding Author:**
Ram Sagar
NASI-Senior Scientist Platinum Jubilee Fellow,
Indian Institute of Astrophysics, Bangalore, 560034
Email: ram_sagar0@yahoo.co.in; ramsagar@iiap.res.in

**Abstract:** A review of modernization and growth of ground based optical and near-infrared astrophysical observational facilities in the globe attributed to the recent technological developments in opto-mechanical, electronics and computer science areas is presented. Hubble Space Telescope (HST) and speckle and adaptive ground based imaging have obtained images better than 0.1 arc sec angular resolution bringing the celestial objects closer to us at least by a factor of 10 during the last two decades. From the light gathering point of view, building of large size (> 5meter aperture) ground based optical and near-infrared telescopes based on latest technology have become economical in recent years. Consequently, in the world, a few 8-10 meter size ground-based optical and near-infrared telescopes are being used for observations of the celestial objects, three 25-40 meter size are under design stage and making of a ~ 100 meter size telescope is under planning stage. In India, the largest sized optical and near-infrared telescope is the modern 3.6-meter located at Devasthal, Nainital. However, the existing Indian moderate size telescopes equipped with modern backend instruments have global importance due to their geographical location. Recently, the Government of India approved India's participation in the Thirty Meter Telescope (TMT) project.

1. **Introduction:** Despite enormous technological developments in the recent years, many physical conditions prevailing in the celestial bodies cannot be produced on the Earth. For example, centers of stars have more than million degree Kelvin temperature at an extremely large gravitational pressure (~ 250 billion times of the gravity on the Earth); matter density has a wide range from a trillion $g/cm^3$ in the case of neutron stars to trillionth part of a $g/cm^3$ in the case of interstellar matter. Mass of the stars ranges from 0.1 to 100 $M_{sun}(=2 \times 10^{33} g)$ while that of galaxies are $\sim 10^{12}$-$10^{13}$ $M_{sun}$. Therefore, their scientific studies, called Astrophysics, provide us with a laboratory in which physical laws and theories are applied, tested and refined at temperatures, pressures and scales otherwise unattainable. Astrophysics is thus an observational science and distinctly different from other branches of laboratory experimental sciences. A schematic diagram for the observations of radiation from a celestial body (star) is depicted in Fig. 1. Before reaching the Earth's surface, the radiation first travels in space and then in the Earth's atmosphere which absorbs almost all the wavelength region of electromagnetic spectrum except optical, near-infrared (IR) and radio windows. Even the electromagnetic radiations passing through the Earth's atmosphere get attenuated as well as blurred.

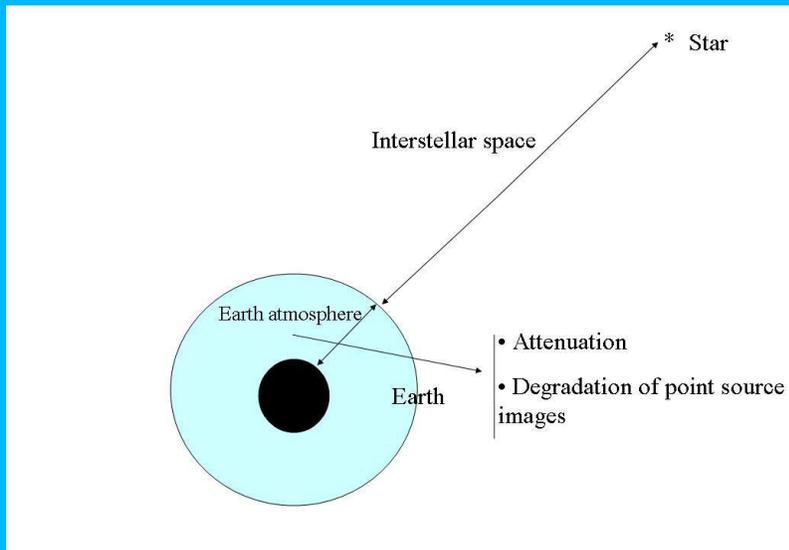

**Fig. 1** Effects of Earth's atmosphere on the light of a celestial body (star) are shown

Historically, front line studies in the area of Astronomy and Astrophysics have made advances in mathematics and physical sciences. During Babylon time, the solution of astronomical problems required new ways of doing mathematics, and in particular led to the development of geometry and trigonometry. Two thousand years later, Newton's interpretation of the observations of planetary positions resulted in the laws of motion and mechanics and inspired the development of calculus. Another three hundred years later, Einstein's study of the Universe gave rise to the general theory of relativity and Eddington's concept of nuclear fusion paved the way to explain the energy source in stars. Now-a-days, relativists, cosmologists and high-energy physicists are testing their theories taking the Universe as a laboratory. The start of $21^{st}$ century is, therefore, witnessing major developments in astrophysical research throughout the world. In the present era of multi-wavelength astronomy, India has world class observing facilities at radio wavelength named the giant metre-wave radio telescope (GMRT) developed and operated by National Centre for Radio Astrophysics (NCRA), Tata Institute of Fundamental Research, Pune [1,2]. Its performance has been appreciated globally. On September 28, 2015, Indian Space Research Organisation (ISRO) successfully launched ASTROSAT which is a dedicated astronomy satellite with unique multi-wavelength capabilities in the hard X-ray, soft X-ray, far-UV, near-UV and visible domain [3] making India as the only country to have multi-wavelength space-based observatory capable of monitoring intensity variations in a broad range of cosmic sources. It is expected to have a mission life of 5-years. The other main Indian institutions involved in its development were Tata Institute of Fundamental Research (TIFR), Mumbai; Indian Institute of Astrophysics (IIA), Bengaluru; Inter-University Centre for Astronomy and Astrophysics (IUCAA), Pune and the Raman Research Institute (RRI), Bengaluru. Initial scientific observations obtained from all the payloads are very encouraging.

In the present paper, importance of optical and near-infrared observations of the celestial bodies and a global perspective of the existing and upcoming modern Indian ground based observational facilities in the field are presented along with their international collaborative and geographical location advantages. India's participation in the Thirty Meter Telescope (TMT) mega project is also briefly described.

**2. Optical and near-IR Astronomy in the present era of multi wavelength Astrophysics:** The celestial objects radiate across the entire electromagnetic spectrum. One would therefore like to observe them at various wavelengths so that their physical and chemical processes and other properties are understood. For example, energy spectrum of radiation from a celestial body resembles a black-body in case of its thermal origin while it is quite different in case of non-thermal origin of radiation. This leads to the present era of multi-wavelength astronomy and astrophysics.

For much of the history, almost all astronomical observations were performed in the visual part of the electromagnetic spectrum initially with our eyes and later the optical telescopes. With the advancement of technology mainly in the fields of opto-mechanical, electronics and computer, larger and larger ground and space telescopes were constructed to probe deep into space for understanding the fundamental issues related to the formation of stars and galaxies [4-6]. Observations in optical, radio and some windows of IR and millimeter are carried out from the ground while those in gamma ray, X-ray, ultraviolet, and most of the IR are possible only from space. The rapid growth in the field of space astronomy is due to the advent of the satellite era in the recent decades, and it is likely to play an increasingly prominent role in practically all the wavebands in the coming decades. The advent of new observational astronomy at wavelengths other than optical and near-IR has enabled discovery of a number of new celestial objects and phenomena, e.g., the possibility of finding a black hole; the discovery of binary X-ray sources, gamma-ray bursts (GRBs) and quasars, etc. However, to establish their identity and meaning in astrophysical terms and also to arrive at more definitive and clearer conclusions, optical and near-IR observations are indispensable, as these wavelength bands contain extra-ordinarily rich concentration of physical diagnostics which have been developed from the accumulation of over a century of observations, associated laboratory experiments and theoretical work, though it covers a narrow wavelength range in the entire electromagnetic spectrum [7-9]. Atomic and molecular transitions of common species that are found in typical astrophysical environments occur at the optical and/or IR wavelengths. The crucial determination of distance, age and chemical abundance of nearly all celestial objects remain firmly rooted in the techniques of optical and IR astronomy due to their unique ability to measure redshifts, wide angle coverage with high angular resolution, and high spectral throughput. For example, distances and other important physical parameters of GRBs could be estimated only when optical and/or near-IR observations of their afterglows could be taken, though the event was discovered in late sixties [8, 9]. All these clearly demonstrate the potential of optical and IR waveband astronomy in the present era of multi-wavelength astrophysics [7].The thrust area of Indian astronomers has, therefore, been to develop and participate in world class optical and near-IR observing facilities considering the fact that good Indian astronomical sites play a crucial complementary role internationally. The important factor here is the geographical locations of Indian observing sites (Longitude ~79° East) which locate them in the middle of about 180 degree wide longitude band having modern astronomical facilities between Canary Islands (~ 20° West) and Eastern Australia (~160° East). Consequently, the observations of transient and variable sources which require 24 hours of continuous monitoring to understand the complex phenomena e.g. pulsation of white dwarfs, but are not possible from either Canary Islands or Australia due to day light and other related reasons, are successfully carried out from the Indian sites. Two examples when this was vital were the discovery of the rings of Uranus [10, 11] and the earliest optical observations of GRB afterglows [8, 9, 12, 13].

**3. Purpose of Astronomical optical and near-IR telescopes:-** An astronomical optical and/or near-IR telescope is used to image celestial objects at its focus in the same way that our eye images the objects on our retina. However, telescopes with aperture size larger than about 6-8 mm (size of our eye pupil) collect more photons due to large area, making it possible to study relatively fainter stars. They also magnify the apparent angular diameter of the celestial objects and thus provide better details due to higher angular resolution. It is well known [7] that, for sky background limited observations, signal-to-noise ratio at a frequency,

$\nu$ is $\alpha \ \sqrt{(A_{eff} \times I_t)/(\Omega_D \times B(\nu))}$,

where $A_{eff}$ is the light gathering power of the telescope of diameter $D$ which includes the losses due to optics and the quantum efficiency of the detector used at the focus of the telescope; $B(\nu)$ is the sky background intensity at frequency $\nu$; $I_t$ is the integration time and $\Omega_D$ is the solid angle formed by the diffraction limited image of a telescope of diameter $D$ at wavelength $\lambda$, is an Airy's disc of size $(1.22\lambda/D)$. At a given astronomical observational site for a fixed $I_t$, power of a telescope is therefore $\alpha$ $(A_{eff}/\Omega_D)$. A 2.5-m telescope with 0.5" circular image is thus equivalent in performance to a 5-m telescope with 1" circular image, if other conditions are similar. This is true only for the space-based telescopes. For ground-based telescopes, however, the image degrades due to turbulence in the Earth's atmosphere measured in terms of an angle called seeing by astronomers. Fig. 2 shows the improvement in the resolving power of the optical telescope with time.

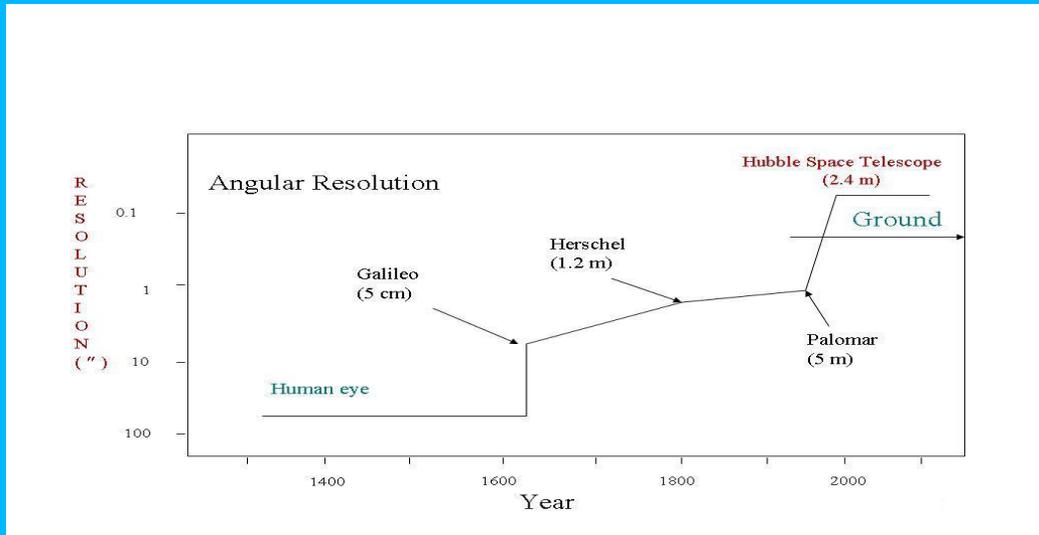

**Fig. 2** Chronological improvement in resolving power of optical telescopes

Most of the recorded seeing measurements indicate that during a large fraction of ground-based observing time, seeing is more than 1" at visual wavelengths. The seeing spreads the image thereby diluting the concentrating action of ground-based optical telescopes of sizes larger than about 15-cm. More turbulence in the Earth's atmosphere at a place, therefore, increases the deterioration due to seeing and hence degrades the image formed by a telescope. This results in much longer hours of observations for recording any faint image or spectra from a ground-based optical and near-IR telescope in comparison to its counterpart in space. For example, Hubble Space Telescope (HST) records images of celestial objects in a relatively shorter time than the ground-based telescopes of similar size. The seeing limited stellar images are much bigger than $\Omega_D$, as is seen from Fig.2. Consequently, significant (an order of magnitude) improvement in angular resolution was observed only twice in the entire history of telescopic observations. First, it was in 1609, when Galileo used a small telescope for observations of stars and planets in place of traditional eye. Second time, it was when images with 2.34-meter HST were taken.

Before describing Indian participation in the world class optical and near-IR observational facilities and future plans, a brief description of the existing Indian optical and near-IR telescopes of aperture size more than a meter is given below.

**4. Existing Indian optical and near-IR Observational facilities:-**India has a 3.6 meter and few 1-2 meter class optical-near IR telescopes. A 4-meter international liquid mirror telescope (ILMT) is in the process of installation. Their location and observing capabilities are given in the following subsections.

I. **Vainu Bappu Observatory, Kavalur:-** Established by IIA in the 1970s, it is located (longitude 78.8° E, latitude 12.6° N and altitude 725 meter) amidst the sandalwood forests in Jawadi Hills in the North Arcot District of Tamil Nadu. There are three telescopes having aperture size larger than 1 meter. They are the 102-cm Carl Zeiss telescope installed in 1972, 234–cm Vainu Bappu Telescope (VBT) installed in 1986 (Fig. 3) [14, 15] and 130-cm telescope installed in 2014. The VBT, a national facility, established in 1986, has f/3.25 prime focus and f/13 Casse grain focus. In order to have a wide field of 20 arc min diameter, Wynne corrector system is used at the prime focus. VBT is equipped with a number of focal plane instruments: both medium and high resolution spectrographs such as fiber-fed Echelle spectrograph, OMR spectrograph and SILFID spectrograph, optical imager, and speckle interferometer.

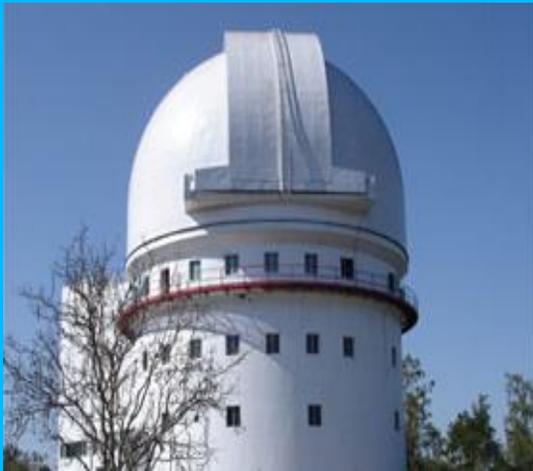

**Fig. 3** A distant view of the 234-cm Vainu Bappu Telescope (VBT)

The 102 cm telescope is equipped with Echelle spectrograph, Zeiss universal astronomical grating spectrograph, polarimeter and CCD imager of different sizes while 130 cm telescope is presently equipped with a CCD imager.

II. **Indian Astronomical Observatory (IAO), Leh-Hanle:** IIA has developed IAO, Hanle in the Changthang trans-Himalayan region of Leh Ladakh district of Jammu and Kashmir. The 2-m Himalayan Chandra Telescope (HCT) was installed at Mt. Saraswati (longitude 75.96 deg E; latitude 32.78 deg N; altitude 4500 m) in August 2000. Figure 4 shows a distant view of the 2-m HCT. It is equipped with state-of-the-art instruments for optical and near-IR imaging and low-resolution spectroscopy. Further details on the telescope and focal plane instruments are given by Prabhu [16].

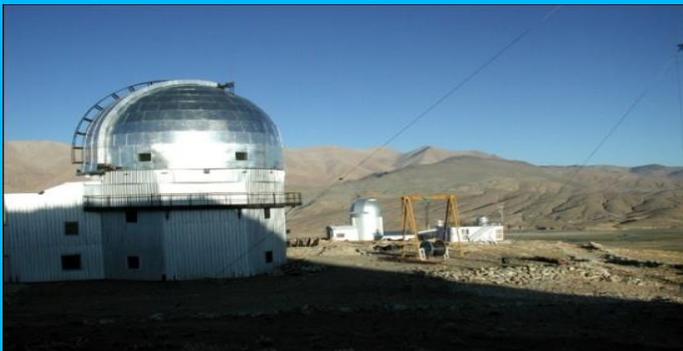

**Fig. 4** A distant view of the 2-meter HCT

III. **Gurushikar Observatory, Mount Abu:-** It houses the 1.2 meter indigenously built optical and near-IR telescope installed in the year 1990 (Fig. 5). The observatory is located at an altitude of 1680 m on the highest peak of the Aravalli range with geographic location of latitude 24.65° N and longitude of 72.78° E and operated by the Physical Research Laboratory (PRL), Ahmedabad. The main focal plane instruments are two channel fast near-IR photometer for lunar occultation; Fibre-linked astronomical grating spectrograph; scanning of IFPS for velocity fields using optical and near-IR emission lines; optical polarimeter; the near infrared camera spectrograph (NICS)and PRL advanced radial-velocity all-sky search (PARAS). Further details of the telescope and focal plane instruments are given by Anandarao and Chakraborty [17]. There is a plan to install a 2.5-meter class optical and near-IR telescope by PRL at the Mount Abu site.

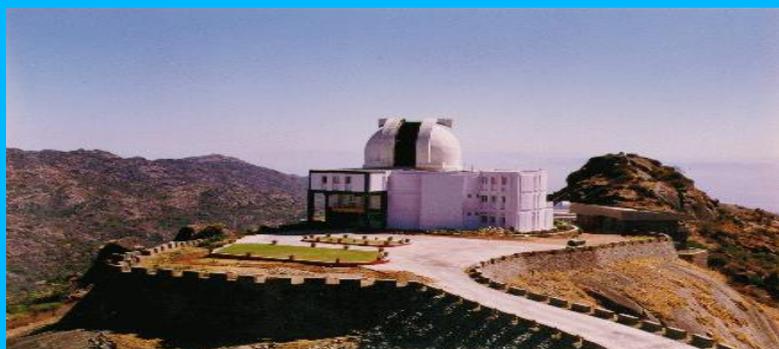

**Fig. 5** A panoramic view of the 1.2 meter optical and near-IR telescope, Mount Abu (reproduced from www.prl.res.in)

IV. **Girawali Observatory:-** The IUCAA installed a 2-meter optical and near-IR telescope on Girawali hill. Its geographical location is latitude 19.07° N; longitude 73.85° E and altitude: ~1000 meters. The telescope was opened for regular observations in November 2006 [18].The telescope is equipped with a faint object spectrograph and TIFR near-IR imaging camera. A distant view of the telescope is shown in Fig. 6.

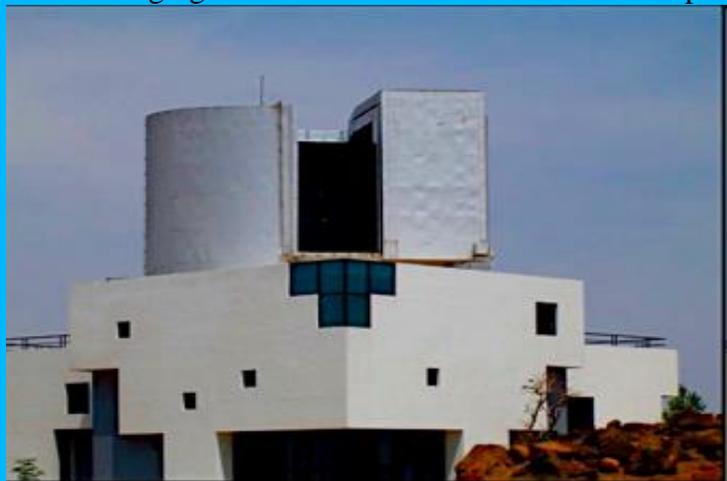

**Fig. 6** A view of the 2 meter IUCAA telescope located at Giravali (reproduced from igo.iucaa.in)

V. **The 104-cm Sampurnanand telescope at Manora Peak**: Geographical coordinates of the telescope at Manora Peak are longitude 79.45° E; latitude 29.36° N and altitude 1951 m. Figure 7 shows a view of the telescope along with a distant view of the ARIES facility. Focal plane of the telescope is equipped with CCD imager of different sizes,

ARIES imaging polarimeter (AIMPOL), photomultiplier tube (PMT) based fast photometer, low resolution spectrograph and near-IR photometer [19].

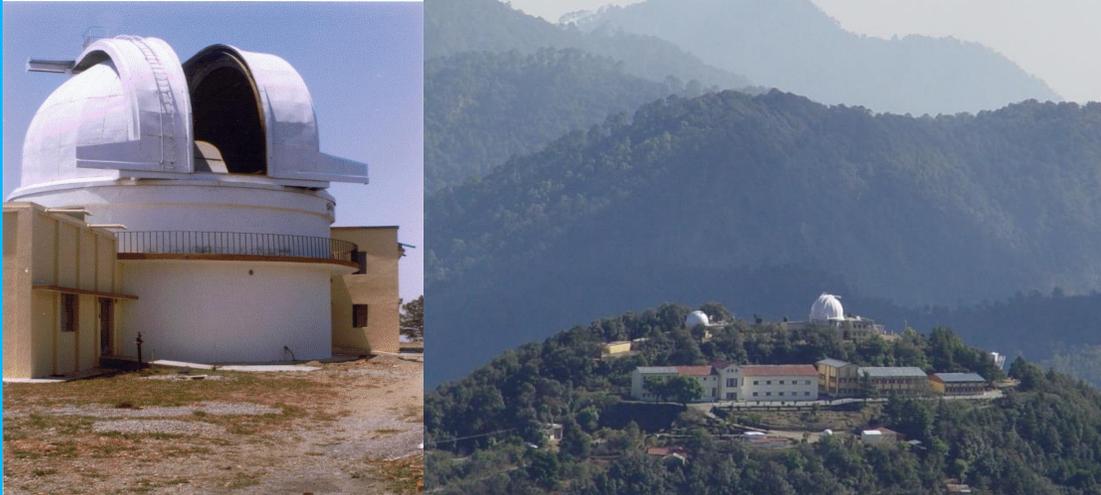

**Fig. 7** A view of (a) the 104-cm Sampurnanand Telescope and (b) a panoramic view of the ARIES facility

**VI. Devasthal observatory:-** Based on extensive characterization of the site [20], ARIES developed the Devasthal site having longitude 79.68° E; latitude 29.36° N and altitude 2450 m, as an observatory for optical and near-IR astronomy. The observatory houses three telescopes of more than 1 meter aperture, namely the 1.3 m Devasthal fast optical telescope (DOFT), 3.6 m dia. Devasthal optical telescope (DOT) and 4-m International liquid mirror telescope (ILMT). Locations and buildings housing these telescopes are shown in Figure 8(a).Technical details as well as observing capabilities of the 1.3m DOFT, installed and dedicated to the nation in 2010, have been described [21-23]. The mechanical structure of the 1.3m DOFT is shown in Fig. 8 (b).

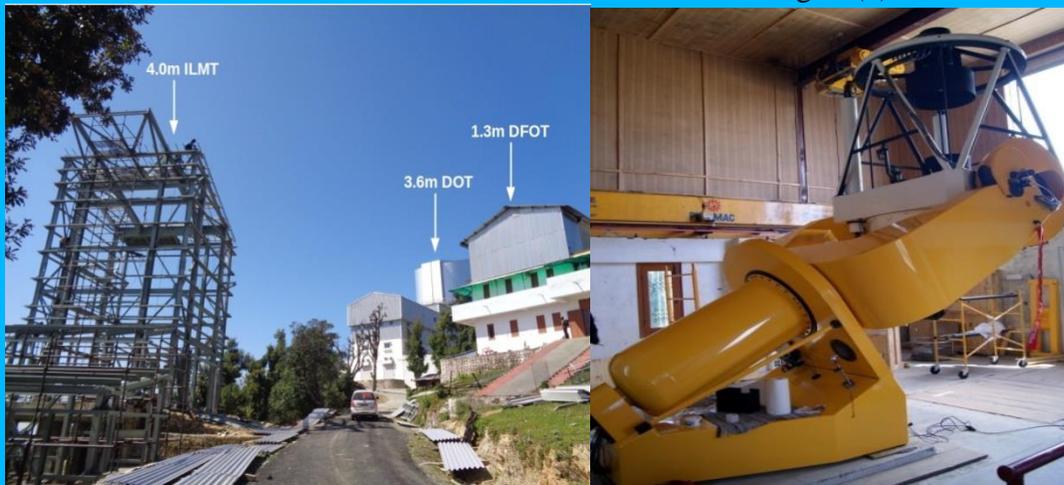

**Fig. 8** (a) Devasthal Observatory showing location of all three telescopes and (b) Mechanical structure of the 1.3m telescope within a sliding dome

India's largest 3.6 meter Devasthal Optical Telescope (DOT) was successfully installed in February 2015 and technically activated on March 30, 2016 (Fig. 9(b)).The f/9, 3.6 m telescope has Ritchey-Chretien configuration with a back focal distance of 2 m. The telescope has an alta-azimuth mounting with a zenith blind spot of less than 5 degree with their technical details [19, 22-24]. The primary mirror of 3.6-m DOT is thin, only 16.5 cm in thickness, which makes it quite flexible under gravity. Therefore, its shape is maintained by active supports pressing against its back under the control of a computer. A relatively thin mirror with active supports was selected mainly due to recent technological progress. This makes the 3.6-m DOT the first new technology telescope in India in which Belgium has contributed 2 million Euros. Such partnerships are extremely valuable for the growth of optical astronomy in India.

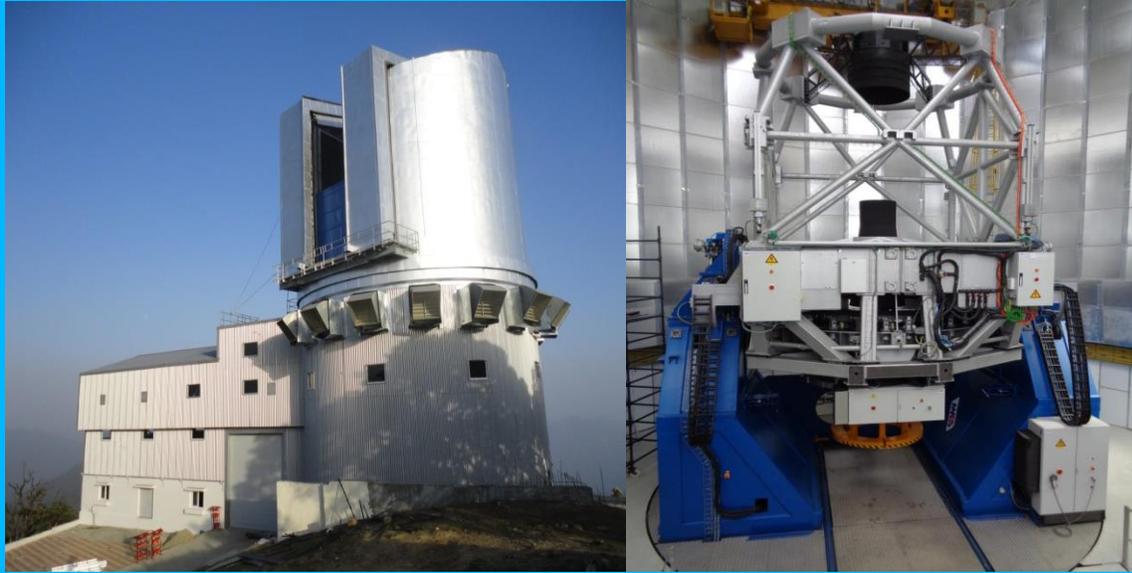

**Fig. 9** (a) Building of the 3.6 meter telescope and *(b)* A fully assembled 3.6m DOT in its dome at Devasthal

A liquid mirror telescope of 4 m size is in the process of being installed at Devasthal. Since it is part of an international effort, it is called the International Liquid Mirror Telescope (ILMT). As the name suggests, the primary mirror of the telescope is a rotating container with highly-reflecting liquid in it. The surface of the spinning liquid (mercury) takes the shape of a paraboloid. Fig 10 shows relative locations of various components of ILMT. Present status of the ILMT building is shown in Fig. 8 (a). The field-of-view of the telescope will be 30' × 30'. ILMT will carry out direct imagery using a 4K x 4K thinned CCD as the detector. At Devasthal, ILMT will monitor a strip of sky of 0.5 degree of declination. This survey will last for about five years. More technical as well as scientific information on ILMT has been given by Poels et al.[25].

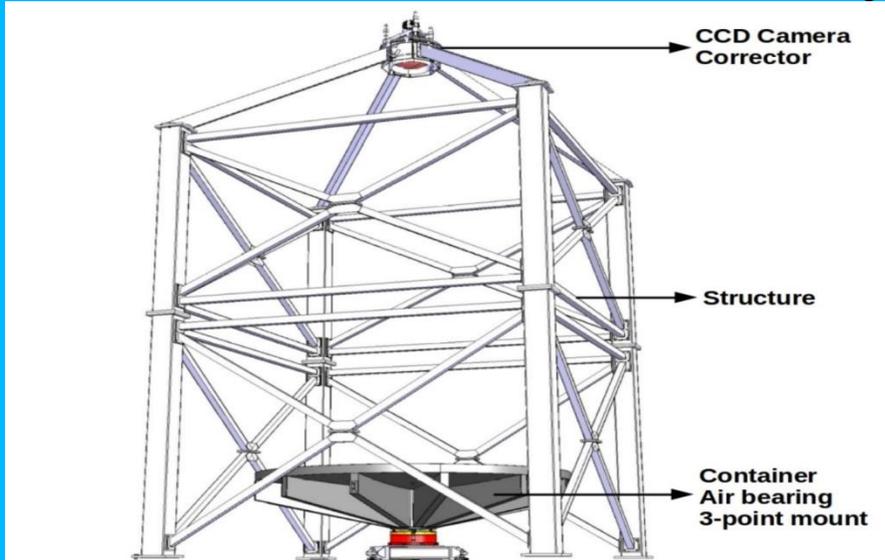

**Fig. 10** A sketch showing the components of the liquid mirror telescope assembled together. The vertical steel frames hold the corrector and CCD detector at the top

**5. Contribution of technological development in building large sized optical and near-IR telescopes:-**In order to understand fundamental science in the Universe such as dark energy and extraterrestrial life etc., astronomers always want to build the largest size telescope feasible by the existing cutting edge technology of that time. However, as can be seen in Fig. 2, by increasing size of the ground based telescopes beyond 15 cm we are not going to improve seeing limited angular resolution of the telescope unless modern techniques of adaptive optics are used [26]. However, light gathering power, $A_{eff}$, of a telescope can be increased by increasing aperture size, decreasing the losses due to optics and increasing the quantum efficiency of the astronomical detectors. Nowadays modern technology is capable of providing optics where losses are minimal (< 1 percent) indicating that there is hardly any scope for further significant improvement in this domain. Impact of modern technology on the astronomical detectors and telescopes are presented in the following subsections.

**5.1 Improvement in modern optical and near-IR detectors and image processing:-** From the invention of the telescope by Galileo in 1600's until late in the 19th century, the detector at the focus of a telescope was the eye which has quantum efficiency (QE) of a few percent only. Then the era of photographic plates/films started. In comparison to the eye, its great advantage was a long exposure time ($I_t$) which built up a picture of a faint object by accumulating more light. However, from the QE point of view, it was similar to the eye. After World War II photo-multiplier tubes having several advantages over photographic emulsions became widely available; for example, unlimited exposure times, larger sensitivity and linear output for measurements of the brightness of astronomical objects could be achieved. The major disadvantage of the photomultiplier tube was that it would observe only a small part of the sky. More recently, astronomical observations started using the technology of television and electronic image amplification with an aim of combining the accuracy and unlimited exposure time of the photomultiplier tube with extended field of view of the photographic plate. The Charge-Coupled Devices (CCDs) are one of such devices. In recent years, photographic emulsions are therefore rarely used for recording astronomical information and in fact, they have become out dated. Nowadays, CCDs are preferred over the other astronomical detectors as they offer a combination of qualities like excellent linearity, high quantum efficiency, large dynamic range, low system noise and dark current, and good overall system stability.

Powerful computers and reduction procedures are as important as modern detectors, and some time even more for extracting the last bit of information as well as to produce the best possible astronomical results from the CCD imaging. The extensive and time consuming numerical analyses used in these data reduction procedures could be carried out due to availability of powerful and cheap computers in recent times. The best examples of this are in doing stellar photometry in crowded regions like globular clusters and estimating completeness of data as a function of brightness in the studies of luminosity functions [27] using both ground and space based CCD images.

The improvement in astronomical detectors and image processing techniques, the availability of highly sensitive spectrographs, the active and adaptive optics, all make a modern optical telescope far more efficient today than its counter-part a couple of decades ago. For example, CCDs in combination with 1-m class optical telescopes are therefore capable of capturing as faint objects as photographic plates can record on a 3-4 m size telescope in a relatively shorter time. All these aforesaid discussions clearly reveal that modern astronomical detectors and image processing techniques with powerful computers have reached a stage where scope of further significant improvement in them may not be seen in the near future. So, to further increase the light gathering power, $A_{eff}$, of the telescope observing system, increasing the diameter of the primary mirror is the only option keeping the ground truth in mind that the largest size of a telescope which can be built at an epoch is mostly decided by the technological and financial limitations. A historical aspect of this aspect has been given in the next subsection.

**5.1.1 Impact of technological development on modern optical and near-IR telescopes:-**
During last few decades, following major developments have taken place mainly in the field of cutting edge technology of modern optical and near-IR telescopes of sizes larger than 3 meter:-

a) **Telescope domes with minimum thermal mass:-** Before 1980s, most of the telescopes had equatorial mounting requiring a steady and constant motion only along their polar axis to compensate rotation of the Earth. However, availability of sophisticated computer control in recent years changed the telescope mounting from equatorial to the much more compact alt-azimuth mount needing non-linear motions of both telescope axes in order to track a celestial body by compensating the rotation of the Earth. This has resulted in reducing the dome size of the telescope building. For example, dome size (~28 meter diameter) for the 2.34 meter VBT with equatorial mounting is much larger than the dome size (diameter of 16.5 meter) for the larger alt-azimuth modern 3.6 meter DOT [22]. Also, construction materials having low thermal mass have been used in modern telescope enclosures in place of concrete structures having larger thermal mass. These along with innovations in dome and building design and environmental control have reduced the Earth's atmospheric turbulence around the telescope and hence improved the seeing and thus the quality of the images taken with modern optical telescopes. Technical details of such improvements in the case of 2-m HCT, 2-m IGO and 3.6-m new technology telescope (NTT) of the European Southern Observatory (ESO) are provided by Prabhu [16], Gupta et al. [18] and Tarenghi and Wilson [28] respectively.

b) **Thin mirror technology** took a quantum leap forward with active control of the shape of the primary mirror and positional control of the secondary mirror together enabling better imaging performance. This was first successfully implemented in the European Southern Observatory's (ESO) 3.6-m new technology telescope in 1988 by Tarenghi and Wilson [28, 29]. The telescopes of sizes larger than 3-meter designed and built during last 2 decades are therefore using active support for both the primary and secondary mirrors. Four 8.2-meter diameter primary mirrors having a thickness of only 2cm have been successfully made and are being routinely used in the very large telescope (VLT) of the ESO. Weight of such thin mirrors (thickness to diameter ratio 1:400) reduces significantly in comparison to primary mirrors having thickness to diameter ratio 1:6 used earlier in order to avoid flexure against gravity while telescopes are pointed towards different direction in the sky. It is mainly because of this reason, that India's largest sized 3.6 m DOT, the first modern technology optical and near-IR telescope of the country, uses the latest technological developments taken in this field [22, 30-36]. Construction of DOT can therefore provide valuable technological knowhow for the young engineers and scientists of the country and go a long way in developing 8-10 meter sized Indian optical and near-IR telescope.

c) Major advancements are taking place in adaptive optics, which effectively control flexible mirrors at mechanical resonance frequencies of 10-20 Hz in order to remove the effects of the blurring of images caused by the turbulence of the Earth's atmosphere [26, 37, 38]. A combination of all these improvements means that a new technology ground-based optical telescope has the capability of approaching the ultimate limit in angular resolution namely the diffraction limit of the telescope itself. This kind of performance is normally associated with telescopes in space, but is now becoming achievable with ground-based optical telescopes at a fraction of the cost of a space mission. In the optical region the gap between the capabilities of space and ground-based telescopes is therefore gradually being reduced.

**6. Extremely large sized optical and near-IR telescopes under construction over the globe and India's participation and future plan:-** Fig. 2 showed that Galileo in 1609 achieved an angular resolution/magnification which was an order of magnitude improvement in comparison

to the naked eye. Since then successive increase in sizes of the optical and near-IR telescope has widened our horizon of knowledge about the Universe. There are a few 8-10 meter class telescopes operational in the globe, namely, twin 10 meter of Keck, Large Binocular Telescope, Subaru and ESO VLT which consists of 4 units, 8.2 m diameter mirrors and 4 movable 1.8 m diameter auxiliary telescopes. These large complex and modern telescopes built during the last 2 decades, equipped with advanced instruments and modern detectors, have provided observational evidence for a number of intriguing questions related to the formation of Universe; e.g. presence of planets around stars and supermassive black holes in the center of many galaxies, distances to the progenitors of powerful gamma ray bursts, evidence of accelerating universe etc. In order to make further front line discoveries in the Universe, astronomical community over the globe aims to move from the present 8-10 meter class telescopes to the extremely large sized (> 20 meter diameter) optical and near-IR telescopes. Building such observing facilities need not only cutting edge and innovative technologies but also huge funds. Institutions and countries across the globe have therefore started forging both financial and technical collaborations. A detailed account of all these has been given recently by Reddy [4]. There are three international teams which are planning to build the next generation of extremely large sized optical and near-IR telescopes that would dramatically dwarf the existing large sized telescopes on the Earth today. They are the Giant Magellan Telescope (GMT), the Thirty Meter Telescope (TMT) and the European Extremely Large Telescope (E-ELT) with corresponding sizes of 25, 30 and 39 m respectively.

During last few decades, India has developed and operating a number of small sized optical and near-IR telescopes(see section 4) but does not have access to any of the above mentioned large sized (8-10 meter) global optical and near-IR observing facilities except the IUCAA's recent acquisition of ~7 % stake in the 10-meter class South African Large Telescope. In fact, the largest Indian facility in the next decade will be the 3.6 meter DOT and 4 meter ILMT. Indian astronomical community therefore decided to join the TMT project after due diligence and thorough deliberations as described by Reddy [4]. In September 2014, Government of India approved our country's participation in the TMT Project with technical and financial contributions amounting to ~10 % of the entire project cost. The TMT project is led by the University of California and California Institute of Technologies (Caltech) in Pasadena, USA with Canada, Japan, China and India [5, 6].The IIA, Bangalore has been identified as the nodal institution for the India-TMT project. Reddy [4] has described the work packages which TMT-India plans to deliver through the participation of Indian companies. In fact, proto type models of a few critical components like actuators and edge sensors have been successfully built by the Indian companies. Active association with the TMT project would help Indian scientists and Engineers to eventually develop and install an Indian 8-10 meter class telescope in the country.

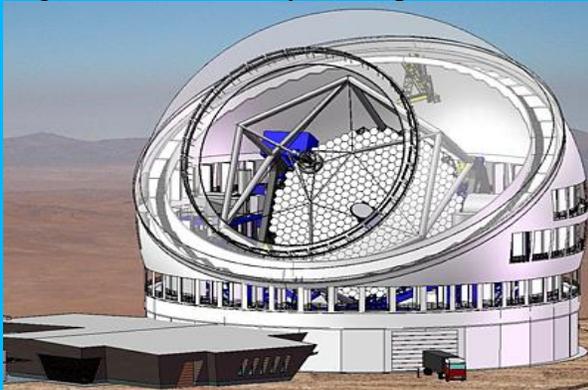

**Fig. 11** TMT building along with primary mirror made of 492 hexagonal segments each of 1.45 m side (reproduced from www.tmt.org)

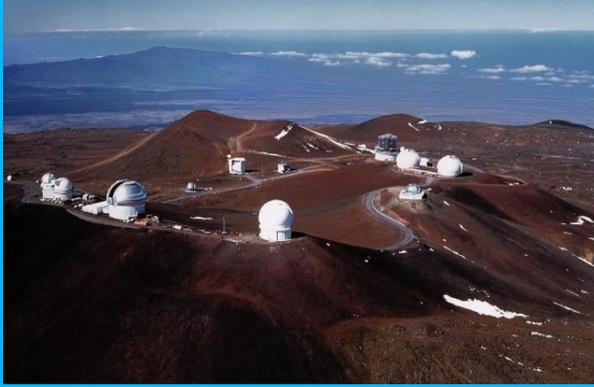

**Fig. 12** Aerial view of the proposed TMT site in Hawaii along with location of the cluster of telescopes like twin Keck, Japanese Subaru and Canada-France-Hawaii Telescope (CFHT) (reproduced from www.tmt.org)

**6. Conclusions:-** Improvement in observing efficiency of an optical and near-IR telescope and its focal plane instruments due to recent technological developments in the fields of opto-mechanical, electronics, computers and detectors have led to significant improvements in the angular resolution capabilities and signal-to-noise ratio of the observed image/spectrum. It is expected that the use of adaptive optics will enable us to obtain diffraction-limited images with ground-based optical telescopes making a vast difference in the faintness of the object that can be recorded with them and also for increasing angular resolution which will benefit many areas of astronomy.

Recent technological developments not only have enabled us to successfully build and operate complex and modern large sized (8-10 meter) operating telescopes like ESO VLT, Keck and Subaru etc., but also have encouraged astronomers to design ground based optical telescopes as large as ~ 40 meter in size and to plan for a ~ 100 meter size telescope as a global venture. Both technological and financial requirements of these mega projects are such that no single country and /or Institute can afford to build and operate them on its own. India's participation in the TMT project will provide a unique opportunity for advancing both science and technology in the country since several key technologies could be transferred to the country which would help us to eventually build our own 8-10 meter class telescope. One can therefore conclude that optical and near-IR Astronomy has a bright future in India and will surely contribute to the growth of pure science which may lead to the technological advancement for the benefit of human beings.

**Acknowledgments**

I thank the National Academy of Sciences, India (NASI) for an award of NASI-Senior Scientist Platinum Jubilee Fellowship and the Directors of both Indian Institute of Astrophysics, Bengaluru and ARIES, Nainital for providing the necessary support. This article is mainly based on the invited talk delivered by me during the 83$^{rd}$ Annual session and Symposium on **Space for human welfare** meeting of NASI held at Goa University, Goa in 2013.

**References**

1. Swarup G, Anathakrishnan S, Kapahi VK, Rao AP, Subrahmanya CR, Kulkarni VK (1991) The Giant Metre-Wave Radio Telescope. Current Science 60: 95-105

2. Swarup G (2015) Major advances in radio astronomy: Some key questions today. Proc Natl Acad Sci India, Sec A, Phys Sci 85(4): 465-481

3. Singh KP, Tandon SN, Agrawal PC, Antia HM, Manchanda RK, Yadav JS, Seetha S,

Ramadevi MC, Rao AR, Bhattacharya D, Paul B, Sreekumar P, Bhattacharyya S, Stewart GC, Hutchings J, Annapurni S, Ghosh SK, Murthy J, Pati A, Rao NK, Stalin CS, Girish V, Sankarasubramanian K, Vadawale S, Bhalerao VB, Dewangan GC, Dedhia DK, Hingar MK, Katoch TB, Kothare AT, Mirza I, Mukerjee K, Shah H, Shah P, Mohan R, Sangal AK, Nagabhusana S, Sriram S, Malkar JP, Sreekumar S, Abbey AF, Hansford GM, Beardmore AP, Sharma MR, Murthy S, Kulkarni R, Meena G, Babu VC, Postma J (2014) ASTROSAT mission. Proc. of SPIE 9144:15 article id. 91441S 15 pp, DOI: 10.1117/12.2062667

4. Reddy BE (2013) India's participation in the thirty-meter telescope project. Journal of Astrophysics & Astronomy 34:87-95
5. Simard L (2013) The thirty meter telescope: Science and Instrumentation for a next generation observatory. Journal of Astrophysics & Astronomy 34: 97-120
6. Sanders GH (2013) The thirty meter telescope (TMT) : An international observatory. Journal of Astrophysics & Astronomy 34: 81-86
7. Sagar R (2000) Importance of small and moderate size optical telescopes. Current Science 78: 1076-1081
8. Misra K, Sagar R (2009) An insight into the progenitors of gamma ray bursts from the optical afterglow observations. Current Science 96:347-356
9. Sagar R, Pandey S B (2012) GRB afterglow observations from ARIES, Nainital and their importance. in Gamma-ray bursts, evolution of massive stars and star formation at high redshifts, ASI Conf. Ser 5: 1–13
10. Bhattacharyya JC, Bappu MKV (1977) Saturn-like ring system around Uranus. Nature 270: 503—506
11. Bhattacharyya JC, Bappu MKV, Mohin S, Mahra HS, Gupta SK (1979) Extended ring system of Uranus. Moon and the Planets 21: 393—404
12. Sagar R, Mohan V, Pandey SB, Pandey AK, Stalin CS, Castro-Tirado AJ (2000) GRB 000301C with peculiar afterglow emission. Bull Astron Soc India (Rapid Communication) 28: 499—513
13. Resmi L, Ishwara-Chandra CH, Castro-Tirado AJ, Bhattacharya D, Rao AP, Bremer M, Pandey SB, Sahu DK, Bhatt BC, Sagar R, Anupama GC, Subramaniam A, Lundgren A, Gorosabel J, Guziy S, de UgartePostigo A, Castro Cerón JM, Wiklind T (2005) Radio, millimeter and optical monitoring of GRB 030329 afterglow: Constraining the double jet model. Astronomy and Astrophysics 440: 477-485
14. Mallik DCV (1998) Twenty five years of observational astronomy at the Indian Institute of Astrophysics. Current Science 74: 735-745.
15. Bhattacharyya JC, Rajan KT (1992) VainuBappu Telescope. Bull. Astron. Soc. India 20: 319-343
16. Prabhu TP (2014) Indian Astronomical Observatory, Leh-Hanle. In Proceedings of the Indian National Science Academy 80:887-912
17. Anandarao BG, Chakraborty A (2010) PRL Mt. Abu Observatory: New initiatives. ASI Conference Series Vol. 1: 211–216
18. Gupta R, Burse M, Das HK, Kohok A, Ramaprakash AN, Engineer S, Tandon SN (2002) IUCAA 2 meter telescope and its first light instrument IFOSC. Bull Astron Soc I ndia 30: 785-790
19. Sagar R, Naja M, Maheswar G, Srivastava AK (2014) Science at High-Altitude sites of ARIES – Astrophysics and Atmospheric Sciences. In Proceedings of the Indian National Science Academy 80:759-790
20. Sagar R, Stalin C S, Pandey A K, Uddin W, Mohan V, Sanwal B B, Gupta S K, Yadav R

K S, Durgapal A K, Joshi S, Kumar B, Gupta A C, Joshi Y C, Srivastava J B, Chaubey U S, Singh M, Pant P, Gupta K G (2000) Evaluation of Devasthal site for optical astronomical observations. Astronomy & Astrophysics Suppl.144: 349-362
21. Sagar R, Omar A, Kumar B, Maheswar G, Pandey S B, Bangia T, Pant J, Shukla V, Yadava S (2011) The new 130-cm optical telescope at Devasthal, Nainital. Current Science101: 1020-1023.
22. Sagar R, Kumar B, Omar A, Pandey AK (2012) New optical telescope projects at Devasthal observatory. in Ground-based and Airborne Telescopes IV, Proc of SPIE 8444 8444T1–12/doi: 10.1117/12.925634
23. Sagar R, Kumar B, Omar A, Joshi YC (2011) New optical telescope projects at Devasthal observatory: 1.3-m installed and 3.6-m upcoming. In recent advances in observational and theoretical studies of star formation, ASI Conf. Ser. 4: 173-180
24. Sagar R (2007) A modern 3.6 meter new technology optical telescope as a major national initiative in astrophysics. National Academy Sciences Letters30: 209–212
25. Poels J, Borra E, Hickson P, Sagar R, Bartczak P, Delchambre L, Finet F, Habraken S, Swings JP, Surdej J (2012) The international Liquid Mirror Telescope (ILMT) as a Variability Time Machine, New Horizons in time-domain astronomy, Proceedings IAU symposium No.285: 394-396.
26. Ellerbroek BL (2013) A status report on the thirty meter telescope adaptive optics program. Journal of Astrophysics & Astronomy 34: 121-139
27. Sagar R, Richtler T (1991) Mass functions of five young Large Magellanic Cloud star clusters. Astronomy and Astrophysics 250: 324-339
28. Tarenghi M, Wilson RN (1989)The ESO NTT (New Technology Telescope): The First Active Optics Telescope.Proc. SPIE 1114, Active Telescope Systems, 302-313 doi:10.1117/12.960835
29. Wilson RN (1989) First Light in the NTT. Messenger 56: 1-5
30. Flebus C, Gabriel E, Lambotte S, Ninane N, Pi'erard M, Rausin F, Schumacher J M (2008) Opto-mechanical design of the 3.6 m Optical Telescope for ARIES in Ground-based and Airborne Telescopes II, Proc. SPIE Volume 7012, article id. 70120A, 12 pp, DOI: 10.1117/12.787888
31. Ninane N, Flebus C, and Kumar B (2012) The 3.6 m Indo-Belgian Devasthal Optical Telescope: general description in Ground-based and Airborne Telescopes IV, Proc. SPIEVolume 8444, article id. 84441V, 11 pp DOI: 10.1117/12.925921
32. Pierard M, Flebus C, Ninane N (2012) The 3.6m Indo-Belgian Devasthal Optical Telescope: the active M1 mirror support in Ground-based and Airborne Telescopes IV,Proc. SPIEVolume 8444, article id. 84444V, 13 pp DOI: 10.1117/12.925946
33. de Ville J, Bastin C, Pierard M (2012) The 3.6 m Indo-Belgian Devasthal Optical Telescope: the hydrostatic azimuth bearing in Ground-based and Airborne Telescopes IV,Proc. SPIEVolume 8444, article id. 84443Z, 11 pp DOI: 10.1117/12.925943
34. Gabriel E, Bastin C, Pierard M (2012) The 3.6 m Indo-Belgian Devasthal Optical Telescope: the control system in Software and Cyber infrastructure for Astronomy II, Proc. SPIE Vol. 8451 article id. 845128, 10 pp, DOI: 10.1117/12.925960
35. Ninane N, Bastin C, de Ville J, Michel F J, Pierard M, Gabriel G, Flebus C, Omar A (2012) The 3.6 m Indo-Belgian Devasthal Optical Telescope: assembly, integration and tests at AMOS in Ground-based and Airborne Telescopes IV, Proc. SPIE Vol. 8444, article id. 84442U, 10 pp, DOI: 10.1117/12.925927
36. Semenov A (2012) Accomplished the task of production of primary and secondary mirrorsof DOT telescopeunder the project ARIES (India, Belgium, Russia): Fabrication features in Modern Technologies in Space and Ground-based Telescopes and Instrumentation II, Proc. SPIE Vol. 8450, article id. 84504R, 14 pp, DOI: 10.1117/12.924645

37. Spyromilio J, Cameron F, D'Odorico S, Kissler-Patig M, Gilmozzi R (2008) Progress on the European Extremely Large Telescope, Messenger 133: 2-8
38. McPherson A, Gilmozzi R, Spyromilio J, Kissler-Patig M, Ramsay S (2012) Recent progress towards the European Extremely Large Telescope (E-ELT), Messenger 148: 2-8